\documentclass[12pt,preprint]{aastex}

\shorttitle{Relativistic Shock Drift Acceleration}
\shortauthors{Matsukiyo et al.}

\begin{document}


\title{Relativistic Electron Shock Drift Acceleration \\
in Low Mach Number Galaxy Cluster Shocks}


\author{S. Matsukiyo}
\affil{Department of Earth System Science and Technology, Kyushu University,\\
    Kasuga, Fukuoka, 816-8580, Japan}
\email{matsukiy@esst.kyushu-u.ac.jp}
\author{Y. Ohira}
\affil{High Energy Acceleration Research Organization,\\
    Tsukuba, Ibaraki, 305-0801, Japan}
\email{ohira@post.kek.jp}
\author{R. Yamazaki}
\affil{Department of Physics and Mathematics, Aoyama Gakuin University,\\
    Sagamihara, Kanagawa, 252-5258, Japan}
\email{ryo@phys.aoyama.ac.jp}
\author{T. Umeda}
\affil{Solar-Terrestrial Environment Laboratory, Nagoya University,\\
    Nagoya, Aichi, 464-8601, Japan}
\email{umeda@stelab.nagoya-u.ac.jp}

%% editorial office after submission.  Svensmark(1998)

\begin{abstract}
An extreme case of electron shock drift acceleration 
in low Mach number collisionless shocks is 
investigated as a plausible mechanism of initial acceleration of 
relativistic electrons in large-scale shocks in galaxy clusters 
where upstream plasma temperature is of the order of 
10 keV and a degree of magnetization is not too small. 
One-dimensional electromagnetic full particle simulations reveal 
that, even though a shock is rather moderate, a part of thermal 
incoming electrons are accelerated and 
reflected through relativistic shock drift acceleration and 
form a local nonthermal population just upstream of the shock. 
The accelerated electrons 
can self-generate local coherent waves and further 
be back-scattered toward the shock by those waves. 
This may be a scenario for the first stage 
of the electron shock acceleration occurring at the large-scale 
shocks in galaxy clusters such as CIZA J2242.8+5301 
which has well defined radio relics.
\end{abstract}

\keywords{
Acceleration of particles ---
Galaxies: clusters: general ---
Plasmas ---
Shock waves ---
Radio continuum: galaxies ---
Relativistic processes
}



%%%%%%%%
\section{Introduction}
%%%%%%%%

In galaxy clusters, the presence of relativistic electrons with 
energies of around GeV has been implied by radio synchrotron 
emissions \citep[e.g.,][]{willson70,govoni04,fuj01}. 
Although some possible acceleration mechanisms for those 
relativistic electrons have been proposed, 
they are still controversial topics \citep[for a recent review, see][]{ferrari08}. 
The diffuse radio emissions in galaxy clusters are categorized into 
several types such as jet robes from active galactic nuclei, 
radio halos located at the center of galaxy clusters,
and radio relics located at the cluster periphery.
In particular, the origin of radio relics has been a major mystery. 
Recent observations suggest that they are associated
with large-scale shocks which are also thought to be possible 
sources of the ultra-high-energy cosmic rays above 
$\approx 10^{18.5}~{\rm eV}$ \citep{kang97,inoue07}. 
%{\bf 
The existence of shocks in the radio relic implies that the high-energy 
electrons are most likely produced through the diffusive 
shock acceleration process \citep{ensslin98,miniati01,gabici03,keshet04}. 
In fact,  radio shells showing the spectral softening due 
to the synchrotron and the inverse Compton coolings are observed 
in CIZA~J2242.8+5301 \citep{wee10}, 
while a temperature jump is found in A3667 \citep{finoguenov10}.
Observations suggest that the temperature is $T\approx 9~{\rm keV}$, 
the magnetic-field strength $B\approx 5~{\mu G}$, 
and the Mach number $M\approx 4.5$ for  CIZA~J2242.8+5301,
and $T\approx2-5~{\rm keV}$, $B\gtrsim 3~{\mu G}$, 
$M\approx 2$, and the number density is $n\approx10^{-5}-10^{-4}$ 
for A3667.
%}

%
%{\bf 
According to the standard theory of structure formation, 
galaxy clusters result from mergers of smaller structures. 
High Mach number shocks ($M>10$) are formed in accreting 
small halos and around the virialized core of the galaxy cluster.
They can accelerate cosmic rays to ultra-high energies
with harder energy spectrum.
On the other hand,
low Mach number shocks ($M<10$) are formed in the virialized 
core. 
During the mergers and accretions, the most 
kinetic energy of the accreting matter is dissipated by relatively 
low Mach number shocks with $M\lesssim4$ \citep{miniati00,ryu03,kang07}. 
These shocks with $M\approx3-4$ are needed in order to explain 
spectral indices of the radio relic \citep{gabici03}.
In fact, the shock velocity in the radio relic of the galaxy 
cluster is of the order of $10^3$~km~s$^{-1}$ and the 
corresponding Mach numbers are about $M\approx2-4.5$ \citep{wee10,finoguenov10}.
%}
%These values are consistent with simulation results of the cosmological
%structure formation \citep{ryu03,kang07}.
The shock velocity is similar to 
what is observed for young supernova remnant (SNR) shocks which 
generate galactic cosmic-ray electrons \citep{koyama95} 
as well as nuclei \citep[e.g.][]{ohira11}. However, the Mach 
number of a galaxy cluster shock is much smaller than 
that of a SNR shock because of high upstream 
temperature as well as the so-called magnetization parameter,
 $\sigma \equiv B^2_1 / 4 \pi n_1 m_e c^2$.
Here, 
$B_1, n_1, m_e$ and $c$ denote upstream magnetic field, plasma 
density, electron rest mass, and the speed of light, respectively. 
%{\bf
Then, for the galaxy clusters, $T\approx 1 - 10~{\rm keV}$ and 
$\sigma \sim 0.1$ with $B_1=3~{\rm \mu G}$ and 
$n_1=10^{-4}~{\rm cm}^{-3}$.
%}
Such small values of Mach numbers are of the same order as 
those of the shocks observed in the heliosphere, i.e. the terrestrial 
bow shock and interplanetary shocks in which electron acceleration 
is rarely observed \citep{shimada99,oka06,oka09}.

The diffusive shock acceleration is one of the most plausible 
acceleration mechanisms of charged particles at a shock front 
\citep[e.g.,][]{blandford87}, where a resultant energy spectrum 
of the high-energy particles obeys a power law. 
For SNR shocks, galactic cosmic rays 
are thought to be accelerated through this process. 
It is not resolved, however, how a part of particles originally 
composing a thermal background plasma is embedded in 
the diffusive shock acceleration which requires pre-existence of 
nonthermal particles possible to flow freely over the shock. 
This is the so-called injection problem on the diffusive shock 
acceleration. In general, electron injection is thought to be more 
difficult than ion injection because of cross-shock potential. 
Nevertheless, relativistic electrons are believed to be 
accelerated in SNR shocks \citep{koyama95,bamba03,pannuti10}. 
Although a variety of theoretical as well as simulation studies 
have revealed that microinstabilities in the shock transition 
region play some crucial roles, the electron injection process 
even in extremely high-Mach number shocks has not yet been 
understood completely \citep[e.g.,][]{pap88,car88,levinson92,
shimada00,dieckmann00,hoshino02,ohira07,
ohira08,umeda08,umeda09,ah07,amano09,amano10,morlino09,riquelme10}. 

In the above context, electron injection in low Mach number 
shocks is inferred to be even harder, since the free energy for 
microinstabilities in the shock transition region should 
be much smaller than that in high Mach number shocks. 
Although a variety 
of microinstabilities get excited in some heliospheric shocks 
\citep[e.g.,][]{wu84a}, 
it is known, for instance, that the saturation level 
of the modified two-stream instability which is one of the most 
plausible instabilities in shocks observed near the earth is 
not so high as that of the Buneman instability in SNR shocks 
\citep{matsu10}. Given this perspective, an injection process 
of electrons in the galaxy cluster shocks is veiled in mystery. 
Indeed, this problem 
has never been addressed so far. 
Here, we take notice that upstream temperature 
of the cluster shocks is much higher than those 
of SNR and heliospheric shocks. As can be seen in the next section, 
the high upstream temperature enables some electrons to be 
accelerated to relativistic energies via the so-called shock 
drift acceleration (SDA). 

In this paper we investigate a limiting case of the electron SDA 
and associated microprocesses which may be responsible for the 
injection of electrons into the diffusive shock acceleration in 
the galaxy cluster shocks. 
In section~2, the SDA process is briefly reviewed and extended to 
an extreme case. Simulation settings and results are shown in section~3. 
Then, discussions and summary are given in section~4.

%%%%%%%%
\section{Shock Drift Acceleration}
%%%%%%%%

%%%%%%%%%%%
\subsection{Some Basic Features}
%%%%%%%%%%%

Shock drift acceleration (SDA) is one of the efficient acceleration mechanisms 
of charged particles in the transition region of an oblique collisionless shock. 
The motion of non-relativistic electrons in this process is analyzed in detail 
previously (eg. \cite{lm84,wu84b,kv89,kvw89,kvb91}). In the so-called normal 
incidence frame (NIF), where upstream flow direction is parallel to a shock 
normal, the acceleration occurs while an electron stays in the shock transition 
region and drifts along the shock surface due to finite gradient of magnetic 
field strength. The electron gains energy since the direction of the drift 
motion is anti-parallel to the motional electric field.

Assuming that the coplanar magnetic field for a one-dimensional shock is 
in the $x-z$ plane and the shock normal is along the $x$-axis leads to 
%eq1
\begin{equation}
 \dot \gamma = {{\bf p} \cdot {\dot {\bf p}} \over m^2_0 c^2 \gamma} 
 = -{e \over m_e c^2} {\bf v} \cdot {\bf E} 
 = -{e \over m_e c^2} \left( - \dot x {\partial \phi 
 \over \partial x} + v_y E_0 \right).
\end{equation}
Here, $\gamma, {\bf p, v}$ and $e$ are the Lorentz factor, momentum, 
velocity and elementary charge, 
${\bf E}$ electric field, and the dots denote time derivative, respectively. 
Furthermore, ${\bf E} = (- \partial \phi / \partial x, E_0, 0)$ has been 
used, where $\phi$ and $E_0$ denote shock potential and 
the motional electric field. This results in 
%eq2
\begin{equation}
 \label{p_ene}
 \dot W \equiv {d \over dt} \left( \gamma - {e \phi \over m_e c^2} \right) = 
 -{e \over m_e c^2} v_y E_0.
\end{equation}

Some incoming electrons gain significant energy during the drift motion 
and are reflected backward from the shock. These reflected electrons become a 
non-thermal component in the upstream plasma frame. The reflection occurs 
due to the magnetic mirror effect. The de Hoffmann-Teller frame 
(HTF) where upstream flow is along the magnetic field is convenient to 
consider motion of a particle, because the motional electric field 
disappears in this frame. The particle velocity in the HTF 
(${\bf v^{\prime}}$) and the NIF (${\bf v}$) are related as 
${\bf v^{\prime}} = {\bf v} + {\bf V_{HT}}$, where ${\bf V_{HT}} = 
({\bf V_1} \times {\bf B_1}) \times {\bf \hat{n}} / ({\bf B_1} \cdot 
{\bf \hat{n}})$ is the de Hoffmann-Teller velocity, ${\bf \hat{n}}$ 
is the unit vector normal 
to the shock, ${\bf V_1}$ and ${\bf B_1}$ are upstream flow velocity 
and the magnetic field in the NIF, respectively. In the HTF the right hand 
side of equation (\ref{p_ene}) becomes zero so that $W^{\prime}$ is 
conserved. Moreover, if the motion of the 
electron is assumed to be adiabatic, magnetic moment, $\mu^{\prime} = 
p^{\prime 2}_{\perp} / 2 m_e B^{\prime}$, is also conserved. 
Here, the prime denotes a quantity measured in the HTF. This assumption 
is valid while the spatial scale of a shock transition region is 
sufficiently larger than the Larmor radius of the electron. From these 
two restrictions, in order for the electron to be mirror reflected, 
the initial pitch angle, $\alpha^{\prime}_1 = \tan^{-1} 
(p^{\prime}_{1\perp} / p^{\prime}_{1\parallel})$, 
should satisfy the following condition (cf. \cite{fel83} 
for nonrelativistic version).
%eq3
\begin{equation}
 \alpha^{\prime}_1 > \sin^{-1} \sqrt{{B^{\prime}_1 \over B^{\prime}_2} 
 {(\gamma^{\prime}_1 + \Delta \Phi^{\prime})^2 - 1 
 \over u^{\prime 2}_1}}
\end{equation}
Here, $B^{\prime}$ denotes magnetic field strength, $u^{\prime} = 
p^{\prime} / m_e c$, $\Delta \Phi^{\prime} = e (\phi^{\prime}_2 - 
\phi^{\prime}_1) / m_e c^2$, and the subscripts 1 and 2 
indicate upstream (initial) and downstream quantities, respectively. 
If there exists an overshoot, however, $B^{\prime}_2$ and 
$\phi^{\prime}_2$ should be defined there. 

The energy gain in the SDA process is obtained by integrating 
equation (\ref{p_ene}) in the NIF, $m_e c^2 W(t) = -e \int E_0 dy$, 
as the energy gained by an electron during its drift in the 
transition region along the direction anti-parallel to the motional 
electric field (for details to \cite{kvw89}). If one moves 
in the HTF, the energy gain is explained by the well known 
Fermi acceleration mechanism \citep{wu84b}. 
If a particle velocity before reflected is written 
as ${\bf v_{pre}}$, ${\bf v^{\prime}_{pre}} = {\bf v_{pre}} + 
{\bf V_{HT}}$. After being reflected, the parallel velocity 
is reversed so that ${\bf v^{\prime}_{post}} = ( - 
v^{\prime}_{\parallel, pre}, {\bf v^{\prime}_{\perp, pre}})$. 
Then, moving back in the NIF results in ${\bf v_{post}} = 
{\bf v^{\prime}_{post}} - {\bf V_{HT}}$. Now, ${\bf V_{HT}} 
= (0, 0, U_1 \tan \Theta_{Bn})$ when ${\bf V_1} = U_1 
{\bf \hat{n}}$ and ${\bf \hat{n}}$ is along the $x$-axis.
Hence, the acceleration occurs mainly in $z$-direction. 
The more $V_{HT}$ increases, the larger energy gain of the 
particle becomes. According to \cite{kv89}, the acceleration time 
through this process is typically of the order of an ion gyro 
period, $\sim \Omega^{-1}_i$, which will also be confirmed in our 
simulation later. For typical intra-cluster magnetic field of 
$\mu G$, this corresponds to about a few 100 seconds, which is of course 
much shorter than any time scales of incoherent loss processes. 

If $\Delta \Phi^{\prime} = 0$ is assumed, $\alpha^{\prime}_1 > 
\alpha^{\prime}_c \equiv \sin^{-1} (B^{\prime}_1 / B^{\prime}_2)^{1/2}$ 
is the necessary condition for an electron to be 
reflected. When a magnetic overshoot is neglected for simplicity, 
$B^{\prime}_2$ or the critical pitch angle called a loss-cone angle, 
$\alpha^{\prime}_c$, can be determined 
from the Rankine-Hugoniot relation by giving a Mach number ($M_A$), 
an upstream plasma beta ($\beta = 8 \pi n_1 (T_e + T_i) / B^2_1$), and 
a shock angle ($\Theta_{Bn}$) which is an angle between the shock normal 
and upstream magnetic field \citep{tk71}. Top two panels in 
Figure \ref{p_space} show regions in $M_A-\Theta_{Bn}$ 
parameter space where some electrons on a velocity shell with a 
radius of $3 v_{te}$ in an upstream plasma frame of the NIF can be 
reflected, where $v_{te}$ is upstream electron thermal velocity. 
In the two panels, for instance, in the black area some electrons on such 
a velocity shell corresponding to upstream electron beta 
($\beta_e = 8 \pi n_1 T_e / B^2_1$) being equal to 0.1 satisfy the 
condition $\alpha^{\prime}_1 > \alpha^{\prime}_c$, where $T_i = T_e$ 
has been assumed to calculate $\alpha^{\prime}_c$. Similarly, in the 
dark (light) gray area some electrons can be additionally reflected 
by assuming $\beta_e = 0.5 (1.5)$. The top (middle) panel 
corresponds to the case with $\sigma 
= 10^{-4} (3^{-2})$. The figure 
confirms that electrons are hardly reflected when $M_A$ and/or 
$\Theta_{Bn}$ become too large. This may be graphically understood 
as follows. Suppose a velocity shell in $v_z - v_x$ space indicated 
as the broken circle in the bottom panel. Its center corresponds to 
the bulk velocity of the upstream plasma in the NIF, ($U_1, 0$), 
and the radius of the shell is $v_r$. Let us consider whether some 
electrons on this velocity shell is mirror reflected or not 
(In this schematic picture relativistic distortion of the shell 
has been neglected.). 
If $V_{HT}$ is added in $v_z$, the center of the shell shifts 
along the thick gray vertical arrow. Now a broken line connecting 
the origin with the new center of the shell is parallel to an 
upstream bulk velocity in the HTF which is along the upstream 
magnetic field, ${\bf B_1}$. Two solid lines symmetrical with 
respect to the broken line denote a loss-cone. Therefore, the electrons 
outside this loss-cone indicated by the black solid arcs can be 
reflected, otherwise transmit. When $U_1$ or a Mach number 
increases, the center of the shell shifts along the broken line and finally 
a whole shell lies inside the loss-cone. In this case no electrons 
are reflected. Similarly, when $\Theta_{Bn}$ becomes large, 
the broken line gets more vertical and the center of the shell 
walks away from the origin. Then the whole shell again enters 
a loss-cone. The possible reflection area indicated in the 
top two panels are defined like this by assuming the shell 
radius $v_r = 3 v_{te}$. The area expands with increasing 
$\beta_e$ as expected. On the other hand, it does not depend 
much on $\sigma$. In particular when $\sigma < 10^{-3}$, 
distributions of the areas are almost exactly the same as the 
top panel. It is mainly due to the light speed limit of a shell 
radius that the areas look contracted for $\sigma = 3^{-2}$ in 
the middle panel, because the increasing $\sigma$ with constant 
$\beta_e$ results in increase of the thermal velocity. 
(The actual calculation is performed 
by taking relativistic effects into account so that the shell 
is given in momentum space with $p_r = 3 p_{te}$.). 
It is speculated that in extremely high Mach 
number shocks like a SNR shock whose Mach 
number is typically $M_A \sim 10^{2-3}$ electron reflection 
hardly occurs unless there are no preheating mechanisms as discussed 
by \cite{ah07,amano10}. On the other hand, electrons can be 
relatively easily reflected when $M_A < 10$, although acceleration 
may be weak because of small $V_{HT}$. In practice, one should 
note that finite shock potential may reduce the reflection rate 
\citep{wu84b}.

%%%%%%%%%%%
\subsection{An Extreme Case}
%%%%%%%%%%%

As mentioned above, in low Mach number shocks ($M_A < 10$) electrons 
are relatively easily reflected, although the resultant acceleration 
is not so efficient in general. However, if there are some electrons on 
the velocity shell with sufficiently large $v_r$, they may be reflected 
even at large $\Theta_{Bn}$ where $V_{HT} \sim O(c)$. The reflected 
electron energy in such a case would become relativistic. 

A condition 
that an upstream electron on a velocity shell with radius $v_{r}$ 
can be reflected is 
$v^{\prime}_{r} > \left( \sqrt{V^2_{HT} + U^2_1} \right)^{\prime} 
 \sin \alpha^{\prime}_c.
$
When $V_{HT} \gg U_1$, this leads to
$v^{\prime}_{r} > V_{HT} \sqrt{B^{\prime}_1 / B^{\prime}_2}$, where 
effects of shock potential has been again neglected.
If $v_r = \eta v_{te}$, the following condition is finally obtained: 
%eq
\begin{equation}
 \eta > \eta_c \equiv {U_1 \tan \Theta_{Bn} \over c} 
 \left( {B_1 / B_2 \over \beta_e \sigma / 2} \right)^{1/2}
 \label{eta}.
\end{equation}
The inequality (\ref{eta}) implies that reflected electrons can 
be present if there are some electrons on a velocity shell with 
$v_r > \eta_c v_{te}$. The reflected electrons will have 
relativistic energies when $V_{HT} = U_1 \tan \Theta_{Bn} \sim O(c)$. 
In this limit $\eta_c \rightarrow \eta_r$, where
\begin{equation}
 \eta_r = \left( {B_1 / B_2 \over \beta_e \sigma / 2} \right)^{1/2}
 = {\sqrt{B_1 / B_2} \over v_{te}/c}.
\end{equation}
In the solar wind or interstellar plasmas $\eta_r$ 
is extremely large, since electron temperature 
is $\sim 100$ eV at the highest and $\sigma 
\sim 10^{-(4-6)}$ ($\beta_e \sim O(1)$). Therefore, satisfaction 
of the above condition for a shock in such an environment as in 
the earth's bow shock or SNR shocks is almost hopeless. However, 
that may be possible if the electron temperature becomes 
$\sim$ a few$-10$ keV or $\sigma \sim 10^{-(1-3)}$ and 
$\beta_e \sim 0.1-10$ as in some 
large-scale shocks in galaxy clusters \citep{fuj07,nak08,wee10}.

%%%%%%%%
\section{1D PIC Simulation}
%%%%%%%%

In this section one-dimensional particle-in-cell (PIC) simulation 
is performed to reproduce relativistic SDA 
of electrons discussed above in a self-consistent manner.

In the simulation a shock is produced by the so-called injection 
or reflecting 
wall method. An upstream magnetized plasma is continuously 
injected from the left-hand boundary. The plasma is reflected 
at the right-hand boundary and mixture of the incoming and the 
reflected plasmas results in a downstream medium. The shock is 
produced at a boundary between the upstream and the downstream 
plasmas. Since the simulation frame is the downstream rest frame, 
the shock propagates in time from right to left. The 
simulation is done in the NIF so that the injection flow is 
parallel to the shock normal which is along the $x$-axis. 
The upstream magnetic field is in the $x-z$ plane. The 
size of a spatial grid is $\Delta x \approx 0.24 \lambda_{De}$ 
where $\lambda_{De}$ denotes the electron Debye length, and 
the number of super-particles per cell is $N_p = 200$ 
for both electrons and ions. Time resolution is $\Delta t = 
0.06875 \omega^{-1}_{pe}$ where $\omega_{pe}$ is the electron plasma 
frequency. Other physical parameters are shown in Table \ref{table1}. 
For Runs A-C, all injection parameters are common except for 
the shock angle, $\Theta_{Bn}$. The injection Alfv\'{e}n Mach 
number is $M_{Ain} = 5$ which results in $M_A \approx 7.8$ or 
$U_1 \approx 0.061 c$ in the shock frame for Run A, for instance, 
where $U_1$ denotes the upstream bulk velocity. This value roughly 
corresponds to the fast Mach number $M_f \approx 2.9$ by using 
a definition of ion acoustic velocity as $C^2_s =(3 T_i + 
T_e) / m_i$. The magnetization parameter is $\sigma = 1/9$ and 
the plasma beta is $\beta = 3 (\beta_e = \beta_i)$ so that 
the upstream electron temperature is $\sim 43$ keV. An 
ion-to-electron mass ratio $m_i / m_e = 1836$ is realistic. 
In the last two columns the maximum value of the magnetic field 
roughly measured in the overshoot relative to the upstream 
value, $B_{os} / B_1$, and associated $\eta_c$ calculated from 
equation (\ref{eta}) are denoted for reference. Note that $\eta_c$ 
may in fact be larger if non-negligible potential effects are 
taken into account \citep{wu84b}. 
Most efficient electron acceleration is observed in Run A as 
expected from that $U_1 \tan \Theta_{Bn}/c$ is the largest, 
although in all cases reflected electrons are seen. 
Run D is performed with more realistic parameters where 
$M_{Ain} (= 3)$ and $\sigma (=1/16)$ are reduced. The results 
of Run D are essentially the same as those of Run A and discussed 
in section 4. In the following, results of Run A are focused.
%
% tabel 1
\begin{table}
\caption{\label{table1}Shock parameters}
\begin{tabular}{cccccccccccc}
\hline
\hline
& $M_{A}$ & $M_f$ & $U_1/c$ & $\sigma$ & \hspace{0.07cm} 
$\beta_e$ = $\beta_i$ & $T_e$ (keV) & $m_i/m_e$ & 
$\Theta_{Bn}$ & $B_{os}/B_1$ & $\eta_c$ \\
\hline
Run A & 7.8 & 2.9 & 0.061 & 1/9 & \hspace{0.07cm} 1.5 & 42.6 & 1836 & 85 & 6 & 0.99\\
Run B & 7.7 & 2.9 & 0.060 & 1/9 & \hspace{0.07cm} 1.5 & 42.6 & 1836 & 80 & 6 & 0.48\\
Run C & 7.4 & 2.8 & 0.058 & 1/9 & \hspace{0.07cm} 1.5 & 42.6 & 1836 & 60 & 6 & 0.14\\
Run D & 5.0 & 1.9 & 0.029 & 1/16 & \hspace{0.07cm} 1.5 & 24.0 & 1836 & 87 & 3.5 & 1.4\\
\hline
\hline
\end{tabular}
\end{table}
%

%%%%%%%%%%%
\subsection{Run A}
%%%%%%%%%%%

\begin{figure}
\epsscale{0.8}
\plotone{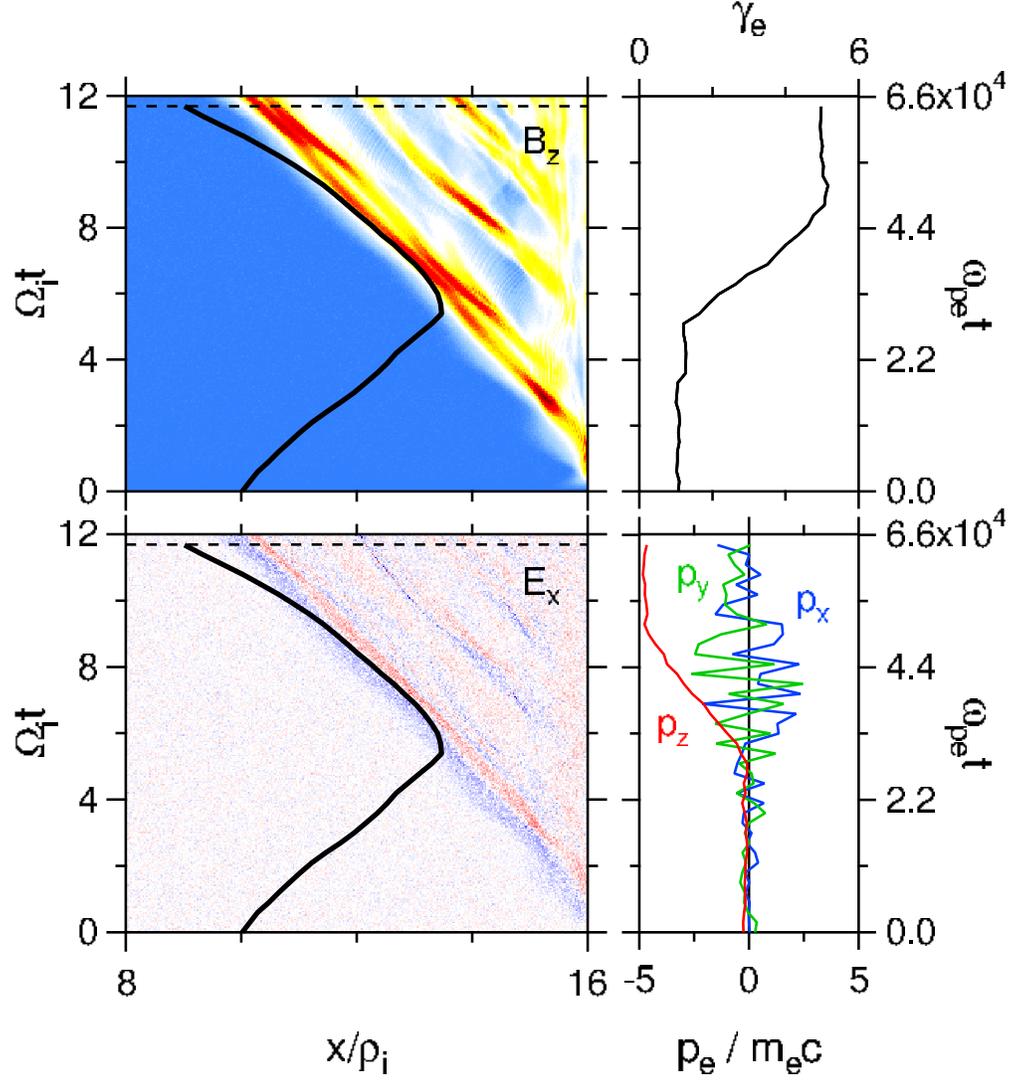}
%\plottwo{f1_color.eps}{f1.eps}
\caption{Left: Spatio-temporal evolutions of color coded $B_z$ 
(upper) and $E_x$ (lower) field components and a trajectory 
of a typical reflected electron. 
Right: Energy (upper) and momentum (lower) histories of the traced 
electron.
\label{XT}}
\end{figure}

In Fig.\ref{XT} spatio-temporal evolutions of $B_z$ and $E_x$ fields 
are shown in the left panels. Structure of the shock is more 
or less time stationary, although it represents weak breathing 
features \citep{com11}. An identical trajectory of 
one of the reflected electrons is denoted as the black solid 
lines. Its energy (upper) and momentum (lower) time histories 
are plotted in the right panels. Here, a rate of output data 
points has been significantly reduced, otherwise $p_x$ (blue) 
and $p_y$ (green) lines fill in the area because of rapid gyro 
motions. The electron gains energy mainly through 
$p_z$ which is almost parallel to the magnetic field during its 
stay in the shock transition region. The time the electron stays 
in the transition region during the reflection process is 
$\sim \Omega^{-1}_i$. These are typical features 
of SDA discussed by \cite{kvw89} and \cite{kv89}, while energy 
of the reflected electron here becomes relativistic.

\begin{figure}
\epsscale{0.5}
\plotone{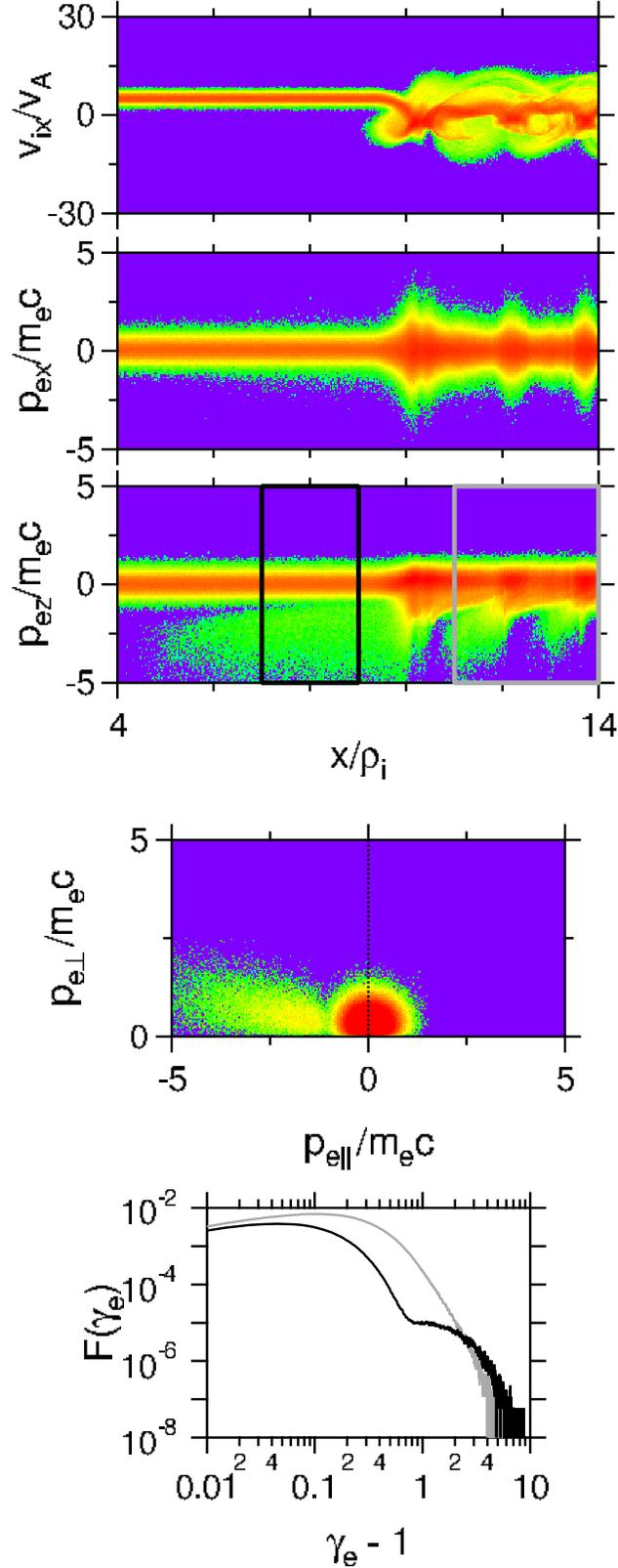}
%\plottwo{f1_color.eps}{f1.eps}
\caption{Phase space densities at $\omega_{pe}t = 64350$. Top three 
panels represent ion $v_x - x$, electron $p_x - x$ and $p_z - x$ 
phase spaces. The fourth and the fifth panels show $p_{\perp} - 
p_{\parallel}$ phase space and a corresponding energy distribution 
functions of electrons surrounded by the squares in the third panel.
\label{phsplt}}
\end{figure}

The first three panels from the top in Fig.\ref{phsplt} show 
ion $v_x-x$, 
electron $p_x-x$, and electron $p_z-x$ phase spaces at 
$\omega_{pe} t = 64350 (\Omega_i t \approx 11.7)$ which 
is indicated as the dashed lines 
in Fig.\ref{XT}. It is clear in the third 
panel that some electrons are reflected basically along 
the magnetic field and have relativistic 
energies. The fourth panel represents $p_{\perp}-p_{\parallel}$ 
phase space of the electrons surrounded by the black square in 
the third panel. Their energy distribution function 
is plotted as the black line in the bottom panel. The reflected 
electrons show a ring-beam feature leading to the non-thermal 
part of the distribution function. A fraction of the bulk energy 
density of the incoming ions carried by the reflected electrons, 
$\epsilon$, is estimated as $\epsilon \approx 0.014$. 
The gray line in the bottom panel is a downstream distribution 
function corresponding to the region surrounded by the gray square 
in the third panel. The high energy part ($\gamma_e - 1 > 1$) 
looks nonthermal. Its origin is the electrons also having 
large negative $p_z$ around $x / \rho_i \sim 11$ and 13 in the 
third panel. They are produced in the second and the third 
magnetic overshoots at $x / \rho_i \sim 12$ and 14 seen in 
Fig.\ref{XT}. By comparing the upstream and the downstream 
energy distribution functions, the highest energy electron is 
upstream. This implies that in the simulation the downstream 
region is still very limited to see an equilibrium state and 
that is clear also from the top three panels. 
%Incidentally, 
%an averaged ratio of ion to electron thermal energies in 
%downstream undershoots is $m_i (\bar{\gamma_i} - 1) / m_e 
%(\bar{\gamma_e} - 1) \sim 5$.

\begin{figure}
\epsscale{0.5}
\plotone{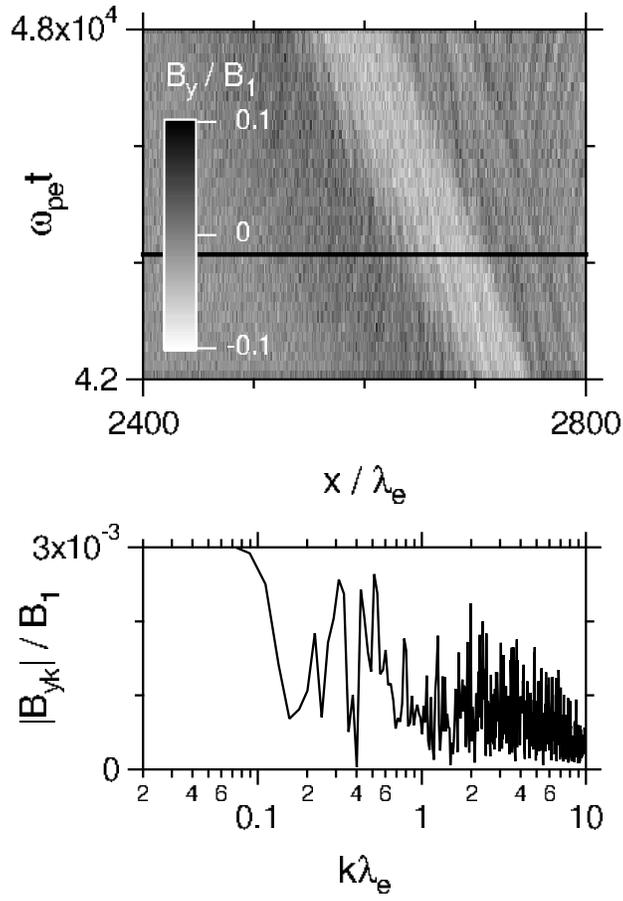}
%\plottwo{f1_color.eps}{f1.eps}
\caption{Small scale fluctuations of $B_y$ in the shock 
transition region (upper panel) and its Fourier spectrum at 
$\omega_{pe}t = 44220$ (lower panel).
\label{TR}}
\end{figure}

It is often thought that reflected 
electrons produce a loss-cone distribution because of 
their adiabatic behaviors \citep{lob05}. 
However, no clear loss-cone is seen in the fourth panel. 
This is probably due to the effects of small-scale waves generated 
in the transition region. Although we avoid getting involved here 
with detailed analysis of this problem, we just show evidence of the 
presence of such small-scale waves. Fig.\ref{TR} shows $B_y$ field 
fluctuations in a transition region (upper panel) and its 
Fourier spectrum at $\omega_{pe}t=44220$ (lower panel) indicated 
by the solid line in the upper panel. 
Tiny streaks in the transition region are visible in the 
upper panel. They appear in the lower panel as spectral peaks 
between $0.1 < k \lambda_e < 1$ where $\lambda_e = c / \omega_{pe}$ 
denotes the electron inertial length. Corresponding structures are 
also seen in ion $v_y-x$ phase space indicating that reflected 
ions destabilize some kind of microinstability (although not 
shown).

%%%%%%%%%%%
\subsection{Injection of Reflected Electrons into Further Acceleration Process}
%%%%%%%%%%%

Here, some possible behaviors of the accelerated electrons after 
being reflected are discussed. Since the reflected electrons 
form the non-equilibrium distribution function, some local 
instabilities may be driven. However, they may not have been properly 
treated in the previous section, because there wave vectors only 
along the $x$-axis are allowed. In the following possible instabilities 
and associated wave-particle interactions are discussed by performing 
further one-dimensional simulations for various wave propagation 
angles with periodic boundary conditions. 
A plasma is assumed to be composed of three components, i.e., 
incoming electrons and ions, and reflected electrons. The reflected 
electrons are assumed to form a ring-beam distribution with a bulk 
momentum $(p_{\parallel0}/m_e c, p_{\perp0} / m_e c) = (2,1)$ and a 
relative density 
$n_b / n_i = 0.1$, where $n_i$ denotes a density of the incoming ions 
which are at rest in average. Hence, the simulation is done in 
the upstream plasma frame of the reference. 
A density and bulk momentum of the 
incoming electrons are decided to satisfy charge and current 
neutral conditions. For all three components, initial temperatures are 
equal and isotropic in their proper frames, $T/m_e c^2 = \beta_e 
\sigma / 2$, where $\beta_e$ and $\sigma$ are chosen to be same 
as the values of the upstream plasma in the previous section 
so that $\beta_e = 1.5$ and $\sigma = 1/9$. 
The system size is $L / \lambda_e = 819.2$, the number of spatial grids 
$N_x = 8192$ corresponding to $\Delta x = 0.1 \lambda_e \approx 
0.35 \lambda_{De}$, the number of super-particles per cell $N_p = 400$, 
and a time step $\omega_{pe} \Delta t = 0.1$, respectively. With 
these parameters fixed, four different runs are performed with 
various wave propagation angles ($\theta_{Bk} = 0^{\circ}, 30^{\circ}, 
60^{\circ}$, and $80^{\circ}$) with respect to the magnetic field 
which is in the $x-z$ plane. Note that $\theta_{Bk}$ is independent 
of the shock angle, $\Theta_{Bn}$.

\begin{figure}
\epsscale{0.5}
\plotone{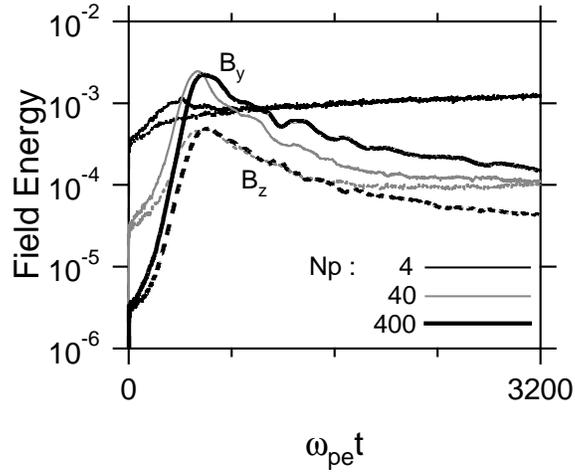}
%\plottwo{f1_color.eps}{f1.eps}
\caption{Time histories of $B_y$ (solid lines) and $B_z$ (broken lines) 
field energies for different numbers of super-particles per cell, $N_p$, 
in case of $\theta_{Bk} = 60^{\circ}$. The black thin lines, the gray 
thin lines, and the black thick lines denote $N_p = 4, 40$, and 400 
cases, respectively.
\label{ene_np}}
\end{figure}

Before discussing simulation results, we first briefly note how the 
number of super-particles per cell, $N_p$, is crucial in reproducing 
a long time evolution of the system. Fig.\ref{ene_np} represents 
evolutions of spatially averaged magnetic field energies as a 
function of time in $\theta_{Bk} = 60^{\circ}$ case for three 
different values of $N_p$; thin black lines correspond to $N_p = 4$, 
thin gray lines to $N_p = 40$, and thick black lines to $N_p = 400$, 
respectively. The solid and the broken lines indicate $B_y$ and $B_z$ 
components. Because of the large noise level, $N_p = 4$ causes 
totally different results from others. For $N_p = 40$, 
early development of an instability appears to be calculated 
properly. However, nonlinear evolution of the system 
($\omega_{pe}t > 1000$) may not be well described. In contrast 
to the case with $N_p = 400$, the $B_z$ component settles in a 
constant value after $\omega_{pe}t = 1600$ and this results in 
the same level of magnitude as the $B_y$ component at last. For 
such an oblique propagation angle, $B_z$ can easily couple with 
electrostatic fluctuations. The lower limit of $B_z$ field 
energy in this nonlinear stage may be an influence of electrostatic 
noise. We confirmed 
that the results are qualitatively common for $N_p \geq 100$ 
at least up to the time to be discussed here. In the following, 
results for $N_p = 400$, which we believe to be reliable, will be 
discussed.

\begin{figure}
\epsscale{0.5}
\plotone{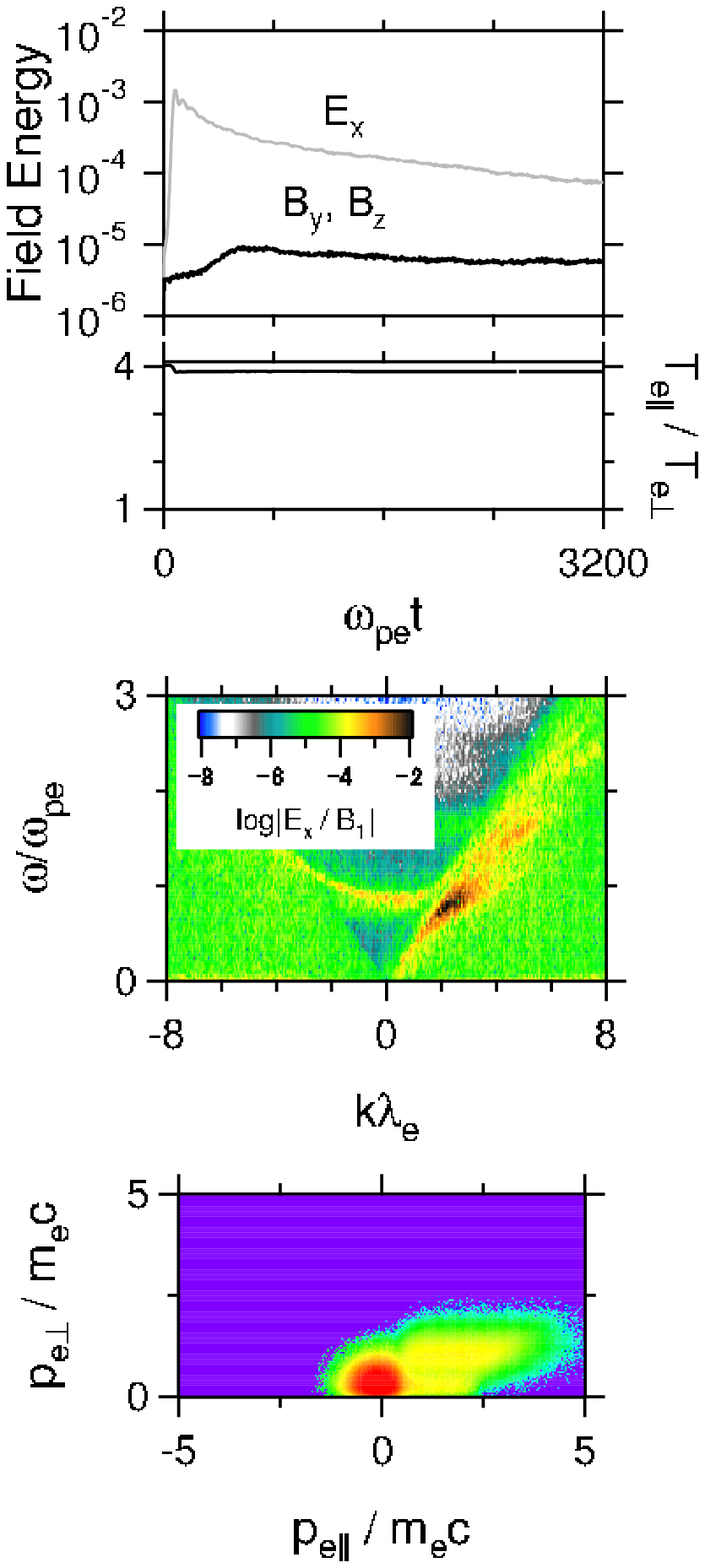}
%\plottwo{f1_color.eps}{f1.eps}
\caption{Time histories of field energies (top) and 
effective electron temperature anisotropy (second), 
$\omega - k$ Fourier spectrum of $E_x$ field for 
$0 < \omega_{pe}t < 204.8$ (third), and a final electron 
distribution in $p_{\perp}-p_{\parallel}$ space (bottom) 
for $\theta_{Bk} = 0^{\circ}$.
\label{th00}}
\end{figure}

Fig.\ref{th00} represents time histories of field energies 
(top) and effective electron temperature anisotropy (second), $\omega-k$ 
spectrum of $E_x$ field for $0 \le \omega_{pe}t \le 204.8$ (third), 
and electron distribution in $p_{\perp}-p_{\parallel}$ phase space 
at $\omega_{pe}t = 3200$, respectively, for $\theta_{Bk} = 0^{\circ}$. 
Here, the effective electron temperature anisotropy is defined as 
$T_{e\parallel}/T_{e\perp} \equiv \Sigma_j n_j p^2_{j\parallel} / 
\Sigma_j n_j p^2_{j\perp}$, where the summation is taken over all 
the incoming and the reflected electrons. 
The rapid growth of $E_x$ field energy is essentially due to 
the beam (two-stream) instability in which the electron 
beam destabilizes mainly Langmuir waves and their higher harmonics 
as shown in the third panel. Although the electron distribution 
function forms a plateau in $p_{\parallel}$ in the end of the run 
as seen in the bottom panel, the temperature anisotropy still 
persists (second panel). Because of the small electromagnetic 
field energies throughout the run ($E_y$ and $E_z$ field energies 
are in the noise level.), the above process is essentially 
electrostatic. Electrostatic wave activities are again dominant 
for $\theta_{Bk} = 30^{\circ}$, although their intensity is much 
less than the case with $\theta_{Bk} = 0^{\circ}$ (not shown).

\begin{figure}
\epsscale{0.5}
\plotone{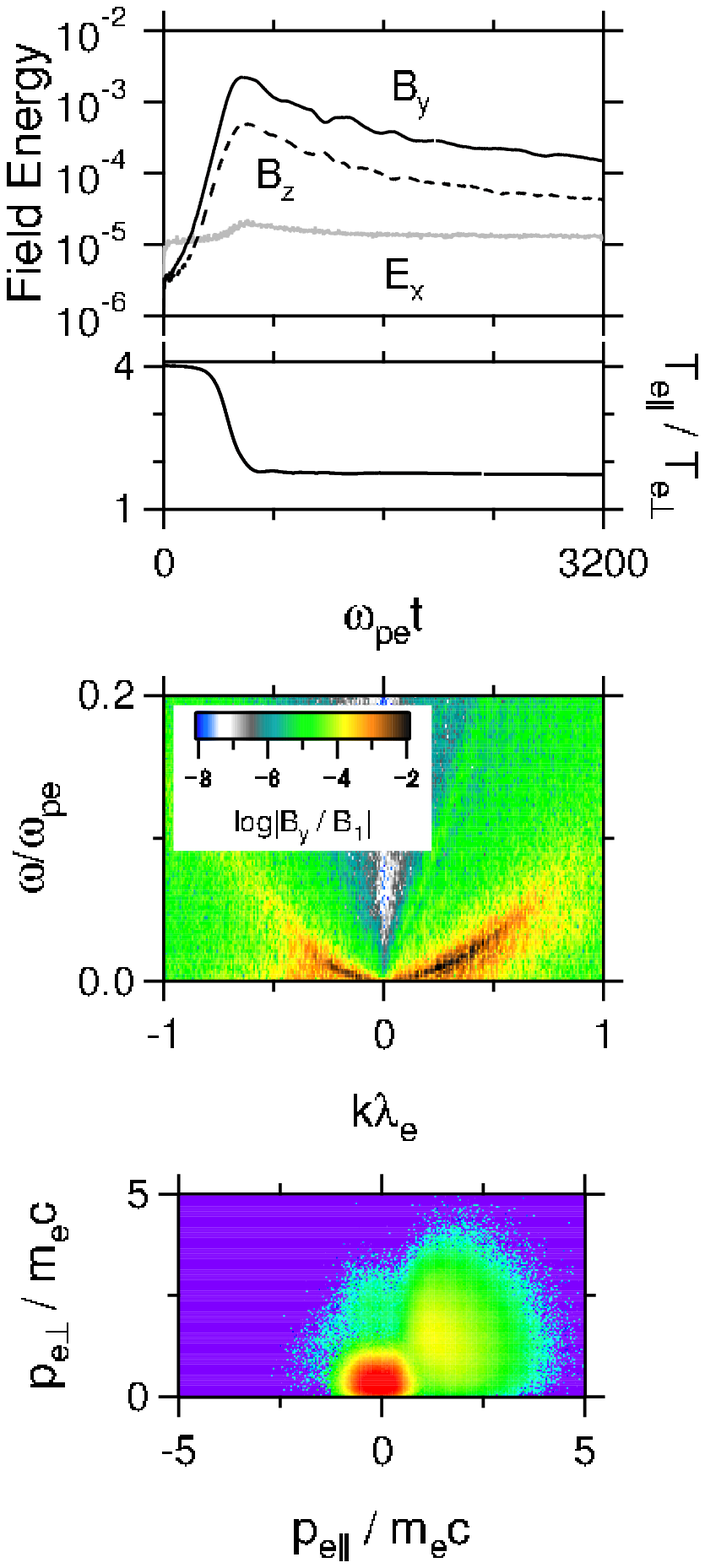}
%\plottwo{f1_color.eps}{f1.eps}
\caption{Time histories of field energies (top) and 
effective electron temperature anisotropy (second), 
$\omega - k$ Fourier spectrum of $B_y$ field for 
$0 < \omega_{pe}t < 3276.8$ (third), and a final electron 
distribution in $p_{\perp}-p_{\parallel}$ space (bottom) 
for $\theta_{Bk} = 60^{\circ}$.
\label{th60}}
\end{figure}

\begin{figure}
\epsscale{0.5}
\plotone{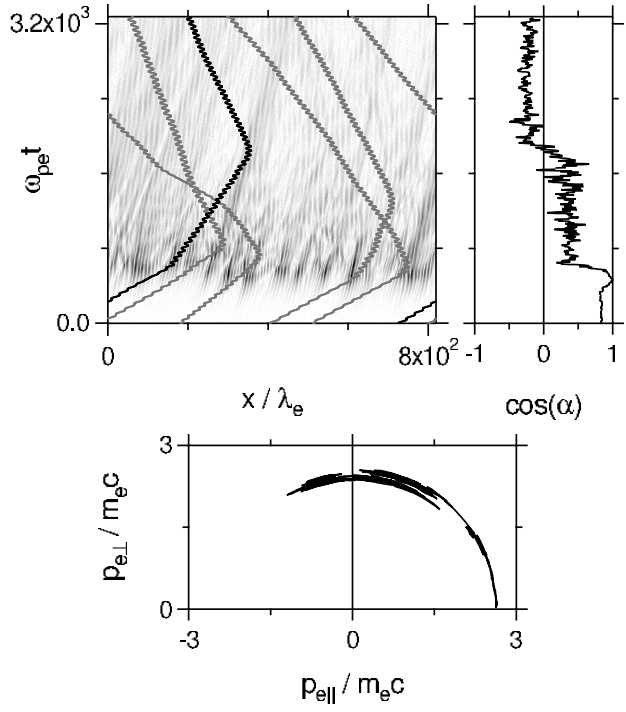}
%\plottwo{f1_color.eps}{f1.eps}
\caption{Trajectories of some electrons back scattered by 
self-generated waves with the background gray scale of 
amplitudes of magnetic fluctuations (top left). 
An evolution of a pitch angle cosine (top right) and a 
trajectory in $p_{\perp} - p_{\parallel}$ space (bottom) 
of the black particle in the top left panel are also shown.
\label{trj60}}
\end{figure}

In contrast, electromagnetic fluctuations become 
predominant for $\theta_{Bk} \ge 60^{\circ}$. The similar plot 
as Fig.\ref{th00} for $\theta_{Bk} = 60^{\circ}$ is shown in 
Fig.\ref{th60}, although the third panel is $\omega-k$ spectrum 
of $B_y$ field for $0 \le \omega_{pe}t \le 3276.8$. This 
electromagnetic instability is essentially nonresonant type 
which is confirmed 
from that natural modes of the beam electrons in the $\omega-k$ spectrum 
(along $\omega \approx k v_{x0}$ ($v_{x0}$ is a projection of a 
parallel bulk velocity into the $x$-direction)) have no 
significant intensities in the third panel. They are probably related with 
one of the oblique modes discussed by \cite{bre09}, although 
detailed linear analysis of kinetic relativistic ring-beam 
instabilities in a magnetized plasma should be reported elsewhere. 
The nonresonant 
instabilities more efficiently relax electron temperature anisotropy 
(second panel) than the electrostatic beam instability. This is 
also confirmed as the widely scattered ring-beam electrons in 
$p_{\perp}-p_{\parallel}$ space in the bottom panel. Especially, 
some of the ring-beam electrons have negative $p_{\parallel}$. 
This means that some of the reflected electrons are back scattered 
by the self-generated waves. (More precisely, electrons having 
$v_{\parallel} \cos \Theta_{Bn} < v_{sh}$ can be regarded as 
being back scattered ($v_{sh}$ denotes the shock velocity), 
although we have not specified here a value of $\Theta_{Bn}$.) 
Some typical trajectories of the back 
scattered electrons are shown as solid lines in the top-left panel 
of Fig.\ref{trj60} in which the background gray scale denotes 
amplitude of the magnetic fluctuation $|{\bf B} - {\bf B_0}|$, where 
${\bf B_0}$ is the ambient magnetic field. 
While all the electrons initially propagate in 
the positive $x$-direction, they finally have negative velocities 
in $x$. Time evolution of the pitch angle cosine of the electron 
corresponding to the black solid line is plotted in the top-right 
panel and its trajectory in $p_{\perp}-p_{\parallel}$ space is 
indicated in the bottom panel. Major changes in the pitch angle 
cosine occur in two bounded time domains where the electron 
encounters large amplitude wave packets ($\omega_{pe}t \sim 500$ 
and $\sim 1800$). In most of the remaining time the electron 
propagates in one-direction, although rapid changes in its pitch 
angle exist. These features are common for all other trajectories 
in the top-left panel.

For $\theta_{Bk} = 80^{\circ}$, a nonresonant instability is again 
dominant (not shown). In such a large $\theta_{Bk}$ generated waves 
are almost nonpropagating pure growing modes which are basic 
features of the Weibel instability. However, growth time is much 
longer and relaxation of the effective electron temperature 
anisotropy is less efficient than $\theta_{Bk} = 60^{\circ}$ 
case.

%%%%%%%%%%%
\section{Summary and Discussions}
%%%%%%%%%%%

It was shown by using a one-dimensional electromagnetic full 
particle simulation that relativistic SDA of electrons is 
feasible even though a quasi-perpendicular shock is rather moderate 
when the upstream electron temperature becomes of the order of 
10 keV or the magnetization parameter $\sigma$ is not too small 
(equation (5)). Such a condition may be realized 
in large-scale shocks of galaxy clusters, cosmic-ray-modified 
subshocks of SNRs, etc. For instance, \cite{wee10} 
showed an evidence of strong acceleration of relativistic 
electrons forming radio relics in the merging galaxy cluster CIZA 
J2242.8+5301, where an average temperature of 
the intra-cluster medium is estimated as $\sim 9$ keV. 
Their polarization analysis indicates that the observed 
shock in the radio relic is quasi-perpendicular. 

The maximum energy of the reflected electrons is affected 
by the de Hoffmann-Teller velocity, $V_{HT} = U_1 \tan\Theta_{Bn}$, 
which is a strong function of a shock angle. The Lorentz 
transformations derive energy of a reflected electron as 
$\gamma_{ref} = \gamma \Gamma^2_{HT} (1 + \beta^2_{HT} - 
2 \beta_{HT} v_z/c)$, where $\gamma$ and $v_z$ denote the electron's 
initial NIF Lorentz factor and velocity component in $z$-direction 
in the shock frame, and $\Gamma_{HT} = (1 - \beta^2_{HT})^{-1/2}$, 
$\beta_{HT} = V_{HT}/c$, respectively. When the nonrelativistic 
limit with large $\Theta_{Bn} (\sim 90^{\circ})$ is considered, 
the above expression reduces to equation (16) in \cite{kvw89}. 
Observed values of the 
maximum Lorentz factors of upstream electrons for Runs A-C 
are $\sim$10.0, $\sim$4.1, and $\sim$2.7, respectively. They roughly 
coincide with the values derived from the above estimate, 
$\sim$9.6, $\sim$3.6, and $\sim$2.3, where 
$\gamma \sim 1.9$ (at a maximum) and $v_z/c \sim -0.8$ 
are again from the simulation. Possible reasons for the 
underestimate may be neglecting a temporal variation of the 
shock velocity and/or nonadiabatic features based on the 
small scale waves in the transition region. 

\begin{figure}
\epsscale{0.6}
\plotone{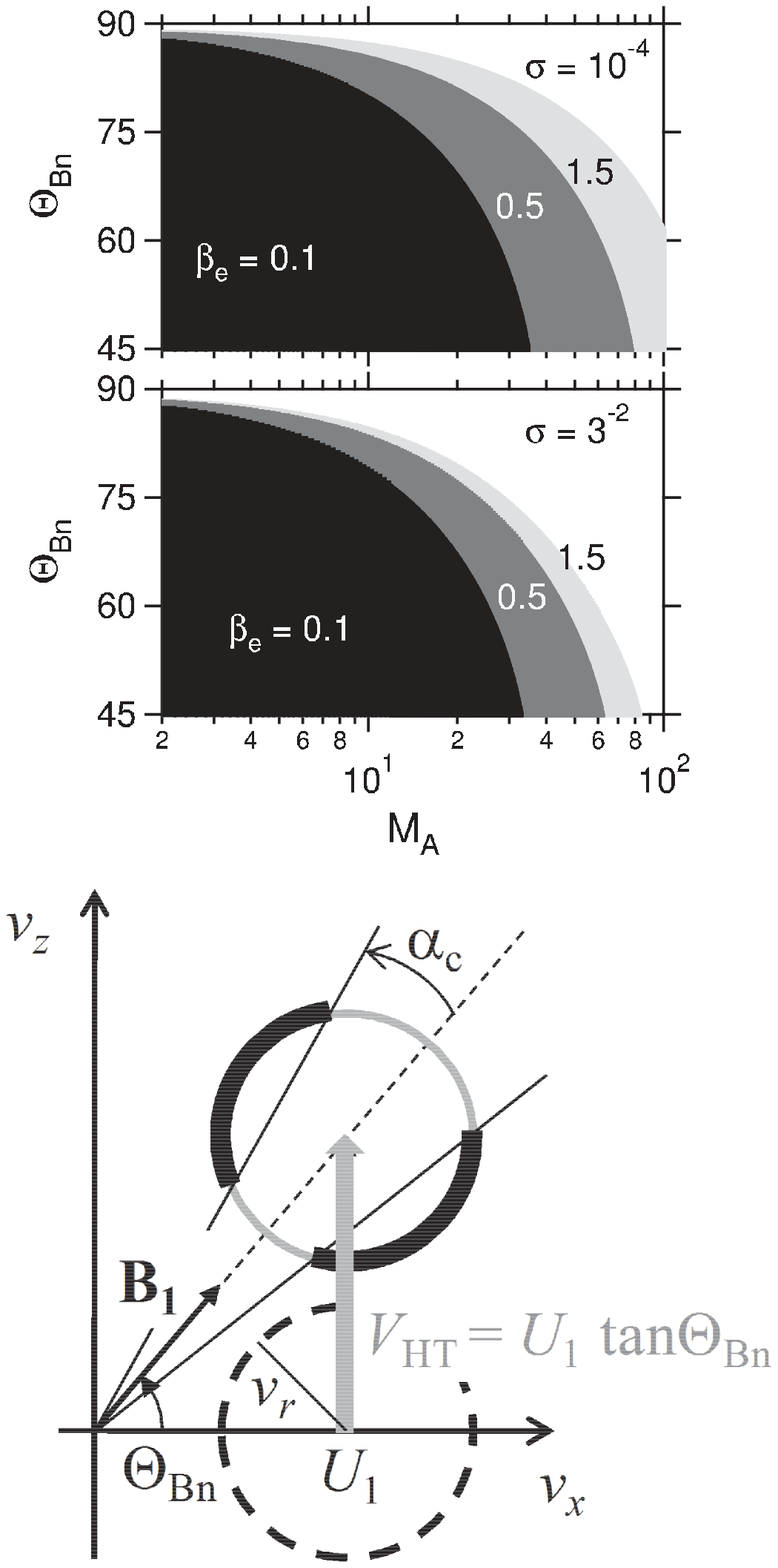}
%\plottwo{f1_color.eps}{f1.eps}
\caption{Feasibility of SDA. Top two panels show $M_A - \Theta_{Bn}$ 
parameter spaces where adiabatic electron reflection is possible for 
three different upstream electron beta values with $\sigma = 10^{-4}$ 
(top panel) and $\sigma = 3^{-2}$ (middle panel). 
The bottom panel shows a schematic of a loss-cone in $v_z - v_x$ space.
\label{p_space}}
\end{figure}

\begin{figure}
\epsscale{1.0}
\plotone{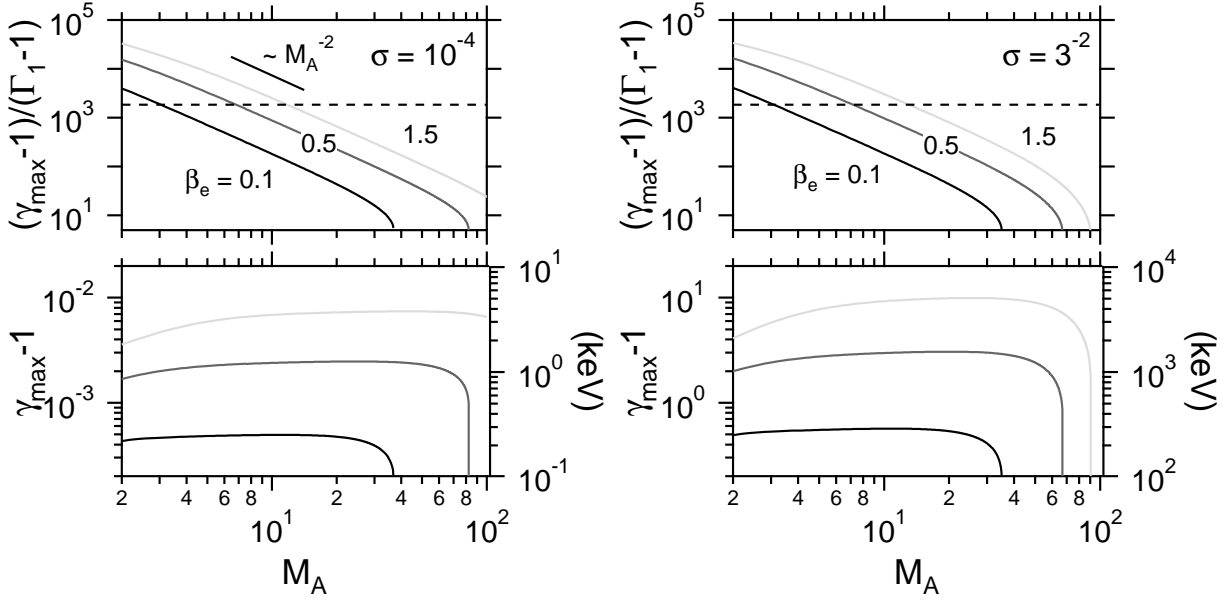}
%\plottwo{f1_color.eps}{f1.eps}
\caption{Parameter dependence of the maximum energy through 
the SDA. The upper panels show the maximum attainable energy 
of the reflected electrons along the upper edges of 
the possible reflection areas indicated in Fig.\ref{p_space} 
normalized to bulk flow energy of the upstream electrons. 
The dashed lines denote bulk energy of the upstream ion flow. 
The lower panels represent the corresponding Lorentz factor 
(left axes) and energy in keV (right axes). The line colors 
indicate different $\beta_e$ (black: 0.1, dark gray: 0.5, 
light gray: 1.5). The left and the right panels correspond 
to $\sigma = 10^{-4}$ and $3^{-2}$, respectively.
\label{Emax}}
\end{figure}

The parameter dependence of the maximum attainable energy 
through the SDA is shown in Fig.\ref{Emax}. Energies of the 
reflected particles along the edges of the possible reflection 
areas in the top two panels 
in Fig.\ref{p_space} are plotted. Each line color corresponds to 
different $\beta_e$ (black: 0.1, dark gray: 0.5, light gray: 1.5). 
The upper panels show energies of the reflected electrons 
normalized to upstream 
electron bulk flow energies. The dashed lines denote bulk 
energy of the ion flow. The normalized energies indicate apparent 
$M^{-2}_A$ dependence for wide range of $M_A$. This is consistent 
with the result that the obtained energies do not basically depend on 
$M_A$ as confirmed in the lower panels. This is due to the 
fact that $V_{HT}$ is roughly constant along the edges of the 
possible reflection area. However, the maximum energy strongly 
depends on $\sigma$ implying that the magnetic 
field strength is crucial. Indeed, it should be 
noted that the vertical axes in the lower right panel is 
three orders of magnitude larger than those in the lower 
left panel, while a ratio of $\sigma$ between these two 
panels is also $\sim 10^3$. 

The reflected electrons are accelerated mainly in $z$-direction 
which is almost along the magnetic field, forming the nonthermal 
part of the energy distribution function upstream of the shock. 
The efficiency of the energy transfer to the reflected electrons is 
highest in Run A. If an injection rate, $\epsilon$, is defined 
as the fraction of energy density carried by the reflected 
electrons with respect to the bulk energy density of the incoming 
ions, $\epsilon \approx 1.4\%$ in this case. The corresponding 
relative number density of the reflected electrons, or a 
reflection ratio, is 2.6\%. 
These values are to be compared with observations. 
\citet{finoguenov10} estimated the total energy of 
radio-emitting electrons in the radio relic in A3667 is $0.15\%$ 
of the shock kinetic energy, 
where they adopt the field strength $B\approx3~\mu$G to explain dim 
inverse Compton emission in the X-ray band.
Another observational estimation of the injection efficiency is
obtained assuming the equipartition between the energies of 
the magnetic field and the radio emitting electrons.
Using the observed quantities given in \citet{finoguenov10},
we derive $4\%$ of the shock kinetic energy goes into
 that of radio emitting electrons.
%
%Although various uncertainties exist, our 
%simulation result on the injection efficiency of $1.4\%$ is 
%comparable to those implied by observations. 
The injection efficiency of a few percent obtained by our simulation result
is comparable to the observational estimation (0.1-4\%),
which, however, contains various uncertainties.
It should be noted that the result here is originated from the 
one-dimensional simulations only for the particular parameter set, 
and that the injection efficiency depends on various parameters such 
as the electron temperature, the shock normal angle, etc. 
To estimate the injection efficiency more accurately, we should perform 
detailed survey on parameters as well as 
two or three dimensional simulations which will be addressed in future works.

The accelerated relativistic 
reflected electrons form a non-equilibrium local ring-beam like 
distribution function upstream of the shock. 
This is possible to generate large amplitude waves through a 
variety of microinstabilities. Local simulations with periodic 
boundary conditions reveal that the electrostatic electron 
beam instability is rapidly destabilized basically along the 
magnetic field. This instability quickly saturates by forming 
a plateau of the distribution function so that the effective 
temperature anisotropy ($T_{e\parallel} / T_{e\perp} > 1$) 
remains. Afterwards, the oblique nonresonant 
instability grows slowly leading to efficient relaxation of 
the temperature anisotropy. In this 
process some of the reflected electrons are scattered back 
toward the shock by self-generated coherent wave packets. 
A sequence of the above process, i.e., the reflection of some 
incoming electrons through the SDA process followed by excitation 
of the microinstabilities and the backscattering of a part of 
the reflected electrons by the self-generated waves, may be 
the first step of injection into the diffusive shock 
acceleration process. %In other words, a production of 
%nonthermal particles and a subsequent injection into the 
%diffusive shock acceleration may seamlessly occur in the 
%large-scale shocks of galaxy clusters.

\begin{figure}
\epsscale{0.5}
\plotone{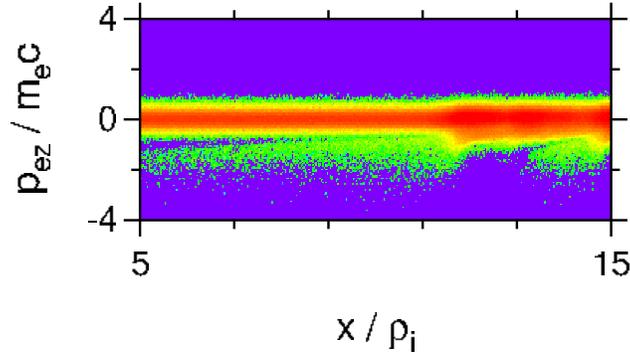}
%\plottwo{f1_color.eps}{f1.eps}
\caption{Electron phase space density in $p_z - x$ at 
$\omega_{pe}t = 85200$ for Run D.
\label{runD}}
\end{figure}

In Runs A, B, and C, $M_A$ and $T_e$ may be somewhat higher 
than those of typical large-scale shocks in galaxy 
clusters. Therefore, an 
additional Run D is performed to confirm that the similar 
process can work also for more realistic parameters. Here, 
$M_A \approx 5.0 (U_1 \approx 0.029c), T_e \approx 24.0$ keV, 
and $\sigma = 0.0625$ as shown in Table \ref{table1}. Results are 
qualitatively similar to Run A. The relativistic SDA works and 
associated ring-beam electrons are produced upstream. However, 
the maximum Lorentz factor of the reflected electrons, 
$\gamma^{max}_{ref} \sim 4$, is not as large as in Run A, since 
$V_{HT}$ and the initial $\gamma$ are a little less because of 
the small $M_A$ and $T_e$ here (Fig.\ref{runD}). 
In addition, a reflection rate of the incoming electrons 
becomes smaller, because $\eta_c \approx 1.4$ which is larger 
than $\eta_c \approx 0.99$ in Run A. Nevertheless, the 
relativistic SDA can work in such a parameter regime too. 
In practice the process may occur even for more moderate and 
lower temperature shocks if a halo electron component is 
present in an intra-cluster medium as in the solar wind 
\citep{stverak09,louarn09}.

In the present paper effects of higher spatial dimensions 
have been excluded. We expect that the processes discussed 
separately in sections 3.1 and 3.2 are simultaneously 
observed in 2D or 3D systems. However, details of the 
competing processes among them have been unknown. 
For example, the electron beam instability is accompanied 
by pitch angle scattering through multidimensional 
wave-particle interactions in its nonlinear stage 
\citep[e.g.][]{pavan09}. If this process proceeds more 
rapidly than the growth of the oblique nonresonant instability, 
the relaxation of the effective temperature anisotropy and 
associated backscattering of a part of the reflected electrons 
might occur through this. It is also curious to know some 
other effects like a rippling and upstream turbulence. All 
these are the future issues.

\acknowledgments

We should like to thank Hiroki~Akamatsu and Susumu~Inoue 
for useful comments and discussions. 
This work was supported by Grant-in-Aid for Scientific 
Research on Innovative Areas 21200050, Grant-in-Aid for 
Scientific Research on Priority Areas 19047004 (R.~Y.), 
and Grant-in-Aid for Young Scientists (B) 21740184 (R.~Y.) 
and 22740323 (S.~M.).
%This work was supported by Grant-in-Aid for Young 
%Scientists (B) No.22740323 (S.~M.), Grant-in-Aid from the 
%Ministry of Education, Culture, Sports, Science, and 
%Technology (MEXT) of Japan, No.~21740184, No.~19047004 (R.~Y.), 
%and Grant-in-Aid for Scientific Research on Innovative Areas 
%21200050. 

%%
%%bibliography
%%

%\clearpage

%figure[fig01bw]
%
%figure[fig02bw]
%
%figure[fig03bw]
%
%figure[fig04bw]
%
%figure[fig05bw]
%
%figure[fig06bw]
%
%figure[fig07bw]
%
%figure[fig08bw]
%
%figure[fig09bw]
%
%figure[fig10bw]

\end{document}